\documentclass[aps,prl,10pt,twocolumn,amsfonts,showpacs,longbibliography,citeautoscript,superscriptaddress]{revtex4-1}

\usepackage{color}
\usepackage{bm}
\usepackage{bbold}
\usepackage{amsmath}
\usepackage{amssymb}
\usepackage{epsfig}
\usepackage{verbatim}
\usepackage[T1]{fontenc}
\usepackage{array}
\usepackage{nicefrac}

\usepackage{tikz}
\usetikzlibrary{calc,arrows,decorations.pathreplacing}
\def\centerarc[#1](#2)(#3:#4:#5){ \draw[#1] ($(#2)+({#5*cos(#3)},{#5*sin(#3)})$) arc (#3:#4:#5); }% Syntax: [draw options] (center) (initial angle:final angle:radius)

\newcommand{\be}{\begin{equation}}
\newcommand{\ee}{\end{equation}}
\newcommand{\bw}{\begin{widetext}}
\newcommand{\ew}{\end{widetext}}
\newcommand{\bea}{\begin{eqnarray}}
\newcommand{\eea}{\end{eqnarray}}
\newcommand{\la}{\langle}
\newcommand{\ra}{\rangle}
\newcommand{\dg}{^\dagger}
\newcommand{\p}{\partial}
\newcommand{\rd}{{\rm d}}
\newcommand{\s}{\sigma}

\def\nn{\nonumber\\}
\def\fr#1{(\ref{#1})}
\def\ocite#1{[\onlinecite{#1}]}

\usepackage[colorlinks=true,bookmarks=false,citecolor=blue,linkcolor=blue,hyperfootnotes=true,urlcolor=blue]{hyperref}

\begin{document}

%%%%%%%%%%%%%%%%%%%%%%%%%%%%%%%%%%%%%%%%%%%%%%%%%%%%%%%%%%%%%%%%%%%
\title{Umklapp scattering as the origin of $T$-linear resistivity in the\\ normal state of high-$T_c$ cuprate superconductors}
%%%%%%%%%%%%%%%%%%%%%%%%%%%%%%%%%%%%%%%%%%%%%%%%%%%%%%%%%%%%%%%%%%%

\author{T. Maurice Rice}
\affiliation{Condensed Matter Physics \& Materials Science Division, Brookhaven National Laboratory, Upton, NY 11973-5000, USA}
\affiliation{Theoretische Physik, ETH Zurich, 8093 Zurich, Switzerland}

\author{Neil J. Robinson}
\affiliation{Condensed Matter Physics \& Materials Science Division, Brookhaven National Laboratory, Upton, NY 11973-5000, USA}
\affiliation{Institute for Theoretical Physics, University of Amsterdam, Science Park 904, 1098 XH Amsterdam, The Netherlands}

\author{Alexei M. Tsvelik}
\affiliation{Condensed Matter Physics \& Materials Science Division, Brookhaven National Laboratory, Upton, NY 11973-5000, USA}

\date{\today}

\begin{abstract}
The high-temperature normal state of the unconventional cuprate superconductors has resistivity linear in temperature $T$, which persists to values well beyond the Mott-Ioffe-Regel upper bound. At low-temperature, within the pseudogap phase, the resistivity is instead quadratic in $T$, as would be expected from Fermi liquid theory. Developing an understanding of these normal phases of the cuprates is crucial to explain the unconventional superconductivity. We present a simple explanation for this behavior, in terms of umklapp scattering of electrons. This fits within the general picture emerging from functional renormalization group calculations that spurred the Yang-Rice-Zhang ansatz: umklapp scattering is at the heart of the behavior in the normal phase.
\end{abstract}

\maketitle

%%%%%%%%%%%%%
{\it Introduction.---}
%%%%%%%%%%%%%
The anomalous temperature and frequency dependence of the electrical d.c. conductivity in the pseudogap (PS) phase of underdoped cuprates has attracted special attention~\cite{GurvitchPRL87,*OrensteinPRB92,*TakagiPRL92,*MandrusPRB92,*AndoPRL01,*DaouNatPhys09,WangPhysicaC96,BarisicPNAS13}. While there is agreement that the PS phase is a precursor to the Mott insulator (MI), a coherent description of its unexpected features is still a challenge. Many of these are in momentum space, e.g. the breakup of the Fermi surface by energy gaps near the antinodes with superconductivity only near the nodal Fermi pockets~\cite{DamascelliRMP03}. At high temperatures the energy gap in the PS phase disappears and the resistivity increases linearly with temperature $T$ to anomalously large values~\cite{BarisicPNAS13}. This radical departure from a standard metal has led to the label of `strange metal' for this phase. In this letter, we argue that the explanation lies in dominant umklapp (U) scattering, i.e. elastic scattering processes that directly transfer momentum between the conduction electron sea and the underlying lattice. 

\begin{figure}
\includegraphics[width=0.3\textwidth]{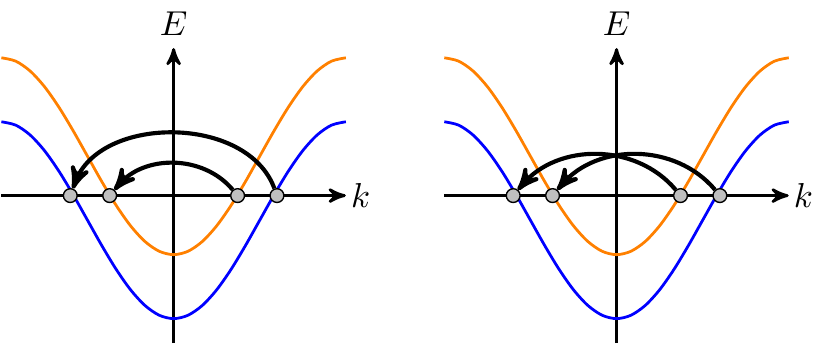}	\\
\includegraphics[width=0.3\textwidth]{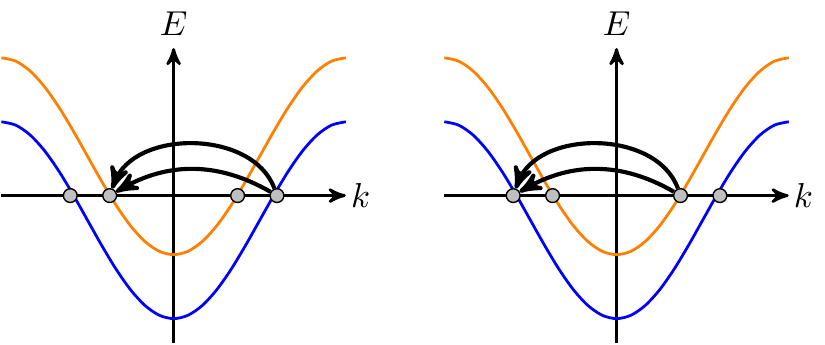}
\caption{Umklapp scattering processes ($RR\to LL$) for right-moving electrons at the Fermi points (grey circles) in the one-dimensional \nicefrac{1}{2}-filled two-leg Hubbard ladder.}
\label{Fig:umklapp}
\end{figure}

The parent undoped cuprates are MIs~\cite{MottProcPhysSocA49,*MottRMP68} -- a state where strong electron-electron interactions cause conduction electrons to condense onto ions, forming an insulating lattice of neutral atoms. In one dimension, however, a Mott state appears already at weak interactions, driven by U-scattering across the Fermi surface~\cite{SolyomAdvPhys79}. This causes momentum transfer between the conduction electron sea and the ionic lattice in units of reciprocal lattice vectors and leads to the insulating ground state. A case of particular interest to us is the \nicefrac{1}{2}-filled two-leg Hubbard ladder (2LHL), and its so-called $d$-MI state, a state that contains seeds of $d$-wave superconductivity~\cite{BalentsPRB96,LinPRB98,*KonikPRB01}. At high temperatures, a one-dimensional (1D) bosonization analysis finds that the resistivity due to umklapp scattering (see the processes in Fig.~\ref{Fig:umklapp}) rises linearly with~$T$, the anomalous form also observed in the `strange metal'~\cite{GiamarchiPRB91,footPhaseSpace}.  

%%% 1D Umklapp scattering figure

The approach that we follow here is to start from the bosonization solution of the 2LHL, and treat the generalization to 2D in the spirit of a $k$-space factorization within a finite patch approximation. This was introduced by Ossadnik~\cite{OssadnikArxiv16}, who argued that short range order yields $k$-space correlations with a finite width, which justifies a wave packet analysis where one breaks $k$-space into finite width patches. When combined with the Yang-Rice-Zhang (YRZ) distortion of the Fermi surface~\cite{YangPRB06,*RiceRepProgPhys12}, this yields a reasonable description of the physics observed within ARPES in high-$T_c$ cuprates~\cite{YangPRL11,*JohnsonJPCS13}, as well as the behavior of the high-field Hall effect within the PS phase~\cite{LiuArxiv17}. Combined with the rather different approach of~\cite{TsvelikPRB17}, these works provide a formal link between the physics of fermionic ladders and 2D doped MIs. The antinodal regions are mapped onto mutually perpendicular \nicefrac{1}{2}-filled ladders, which interact only weakly with the nodal regions (to first approximation, the two regions can be treated as decoupled). In the present letter, we show that this picture captures the behavior of transport within the PS phase of the high-$T_c$ cuprates, with the correct form of the longitudinal transport $\varrho(T)$ being found, as well as explaining the Hall angle as a function of temperature.

\begin{figure}
\includegraphics[width=0.25\textwidth]{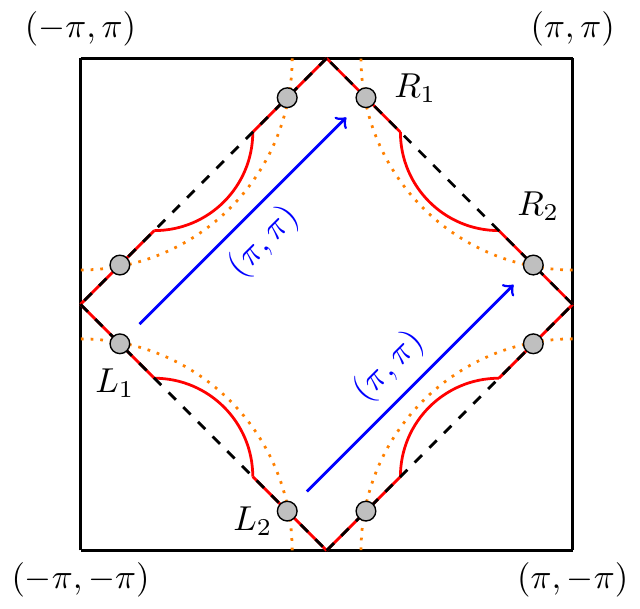}
\caption{The non-interacting Fermi surface (dotted) intersects the umklapp surface (dashed lines, connecting the antinodes) at eight isolated points within the Brillouin zone, known as hotspots (gray circles). In the presence of strong interactions the Fermi surface may deform in order to increasing the nesting (depicted in red), leading to the formation of the well-known `pockets' close to the nodes. One umklapp scattering process, $L_1,L_2 \to R_1,R_2$,  is illustrated.}
\label{Fig:FS}
\end{figure}

%%%%%%%%%%%%%
{\it Umklapp scattering in doped cuprates.---}
%%%%%%%%%%%%%
The importance of U-scattering in the high-$T_c$ cuprate superconductors away from \nicefrac{1}{2}-filling has been emphasized for some time now, see e.g.~\textcite{HonerkampPRB01}. The 2D Hubbard model was studied within the functional renormalization group (FRG) framework; at small dopings away from \nicefrac{1}{2}-filling U-scattering flows to strong coupling. Thus any electrons on the ``umklapp surface'' (the square surface joining the antinodal points, see Fig.~\ref{Fig:FS}) experience strong U-scattering. On the Fermi surface, eight such isolated points exist, known as hotspots. The FRG analysis also showed clear hallmarks of $d$-wave superconductivity and antiferromagnetism~\cite{HonerkampPRB01,HonerkampEPJB02}, in agreement with  exact diagonalization studies~\cite{LauchliPRL04}.

%% 2D Fermi Surface Picture

The presence of the U-surface, on which scattering is particular strong, motivated Yang, Rice and Zhang to propose their phenomenological ansatz for the Green's function in the PS phase~\cite{YangPRB06,*RiceRepProgPhys12}. This puts the U-surface at the center of the physics, with the pairing gap opening on this surface, rather than the Fermi surface. Strong U-scattering then increases the energy gap. The ansatz has successfully described a variety of experiments, from angle-resolved photoemission spectroscopy~\cite{YangPRL11,*JohnsonJPCS13} to the change in the Hall effect upon entering the PS phase~\cite{StoreyEPL16}, and resonant inelastic x-ray scattering~\cite{JamesPRB12,*DeanPRL13,*ShiArxiv16}. 

The important role of U-scattering in the cuprates was again emphasized recently by \textcite{LiuArxiv17,WuPRB17}. They argued that U-scattering causes the emergence of the PS state, turning the superconducting gap at overdoping into an insulating PS at lower doping  due to gaps in both single particle and pair spectra~\footnote{We note that both Refs.~\ocite{LiuArxiv17} and~\ocite{WuPRB17} use concrete realistic parameters in their studies. In the case of Ref.~\ocite{LiuArxiv17}, the focus is on a weaker coupling scenario with $U=0.4t$, $J=0.1t$, $t'=0$. On the other hand, Ref.~\ocite{WuPRB17} considers stronger coupling, of direct relevance to the cuprates, with $U = 5.6t$ and $t' =-0.3t$.}. The presence of strong U-interactions can deform the Fermi surface to stabilize commensurate nesting, which in turn maximizes the interactions, leading to enhanced gaps (similar to Cr alloys~\cite{RicePRB70}) on the YRZ Fermi surface, see Fig.~\ref{Fig:FS} and Refs.~\cite{LiuArxiv17,TsvelikPRB17,WuPRB17}. This deformation of the Fermi surface to run along the U-surface allows a straightforward generalization of the 1D 2LHL analysis, leading to $T$-linear resistivity, to 2D.

%%%%%%%%%%%%%
{\it The normal phases of the cuprates.---}
%%%%%%%%%%%%%
The normal phases  of the cuprate superconductors are enigmatic and enduring mysteries in contemporary condensed matter physics~\cite{DamascelliRMP03,RiceRepProgPhys12,AlloulPRL88,*TimuskRepProgPhys99,*TallonPhysicaC01,*LeeRMP06,*KordyukLowTempPhys15,*ChowduryBook15}. A theory of these phases is crucial to understand the mechanism of high-$T_c$ superconductivity, as these are the states from which superconductivity emerges. The `strange metal' exhibits linear resistivity $\varrho(T) \propto T$, in striking contrast to the conventional $\varrho(T) \propto T^2$ behavior of Fermi liquid theory~\cite{GurvitchPRL87,OrensteinPRB92,TakagiPRL92,MandrusPRB92,WangPhysicaC96,AndoPRL01,DaouNatPhys09,BarisicPNAS13}. Whilst the resistivity may saturate at sufficiently high temperature~\cite{WangPhysicaC96,CalandraEPL03}, it is clear that it can largely violate the Mott-Ioffe-Regel limit~\footnote{The Mott-Ioffe-Regel limit~\cite{IoffeProgSemicond60,*MottPhilosMag72,*GurvitchPRB81,*GurvitchPRB83} corresponds to setting the mean free path equal to the interatomic spacing.}. On the other hand, the PS phase shows the conventional $\varrho(T) \propto T^2$ behavior at low-temperature, with a gradual crossover to linear at the transition to the `strange metal', $T^\ast$~\cite{BarisicPNAS13}. 

The transition temperature $T^\ast$ between the `strange metal' and PS phases is still the subject of much study. The order parameter characterizing the transition, if it exists, is hotly debated. Recent optical studies found a change in symmetry on passing through $T^\ast$~\cite{ZhaoNaturePhys17}. Angle-resolved photo emission spectroscopy~\cite{ZakiUnpublished17}, tunneling spectroscopy~\cite{FischerRMP07}, and Raman~\cite{SacutoRepProgPhys13} studies lend support to the suggestion $T^\ast$ can be related to the PS $\Delta_{PG}$ analogously to the superconducting gap $\Delta_{BCS}$ and $T_c$ in a BCS superconductor, $2\Delta_{PG} \simeq 4.3k_B T^\ast$. 

%%%%%%%%%%%%%
{\it The two fluid model.---}
%%%%%%%%%%%%%
The results of \cite{TsvelikPRB17,LiuArxiv17} present a picture of the PS phase where antinodal and nodal states are well separated. This sets the stage for the \textit{two fluid model} of the conductivity $\s(T)$ in the normal phases~\footnote{A similar model has been used to fit experimental data in Ref.~\cite{ClayholdJSNM10}. Microscopic considerations, such as the origin of the high-temperature $1/T$ and low-temperature $1/T^2$ contributions, did not form part of that work.}
\be
\s(T) \approx \s_\text{nodes}(T) + \s_\text{antinodes}(T), \label{sigma}
\ee
where each contribution comes from nodal or antinodal states. We will see that U-scattering of electrons in the antinodal regions at high temperatures leads to $\s_\text{antinodes}(T) \sim 1/T$ (i.e., $\varrho(T) \propto T$), while conventional scattering around the nodal regions leads to the Fermi liquid form $\s_\text{nodes}(T) \sim 1/T^2$ (i.e., $\varrho(T) \propto T^2$). 

%%%%%%%%%%%%%
{\it Linear resistivity for $T>T^\ast$.---}
%%%%%%%%%%%%%
At the center of our argument for linear resistivity $\varrho(T)\propto T$ at high temperature is the mapping of electrons along the U-surface (i.e., in the antinodal region and the vicinity of the hotspots) to effective ladder models, see Ref.~\cite{TsvelikPRB17,LiuArxiv17}. U-scattering of such electrons dominates the resistivity~\cite{supplemental}, both due to the propensity of U-scattering to dissipate momentum and the strength of such terms (which flow to strong coupling under the FRG~\cite{HonerkampPRB01}).  Linear resistivity then follows in a straightforward way; at an elementary level, this can be seen from the na\"ive Boltzmann argument presented earlier. At a more quantitative level, we can consider the resistivity generated by presence of a U-term $H_U = \int \rd x\, {\cal H}_U$ in the low-energy effective theory~\cite{BalentsPRB96,GiamarchiBook,supplemental} (for details we refer the reader to \cite{TsvelikPRB17}), 
\be
{\cal H}_U = u \Big[ \cos(\sqrt{8\pi}\Phi_{1,c}) + \cos(\sqrt{8\pi}\Phi_{2,c})\Big], \label{umklapp}
\ee
where $u$ is the U-interaction strength and $\Phi_{1,c}$, $\Phi_{2,c}$ are charge fields in the low-energy description of the ladder. Their two-point correlation function is
\bea
&&\left\la \Phi_{a,c}(x,t) \Phi_{a,c}(0,0)\right\ra \label{2pf} \\
&& = \frac{K_c}{4\pi}\ln \left[\frac{\pi T}{\sinh\Big( \pi T (t - x - i 0) \Big)} \frac{\pi T}{\sinh\Big(\pi T (t+x - i0)\Big)}\right], 
\nonumber 
\eea
where $K_c$ is the Luttinger parameter for the charge degrees of freedom~\cite{GiamarchiBook}. 

The charge field in the the presence of a constant current $J$, denoted $\tilde \Phi_{i,c}$, can be related to the zero current field through 
$\Phi_{i,c}(x,t) = \tilde \Phi_{i,c}(x,t) - \pi J t.$
The resistivity can then be computed following the approach of Ref.~\cite{AtzmonPRB12}. The voltage drop $V$ across the system is related to the time-derivative of the dual charge fields $\Theta_{a,c}(x,t)$
\be
\begin{split}
 eV  &= \sum_{a=1,2}\left\la \frac{\rd}{\rd t}\left[\Theta_{a,c}\left(\frac{L}{2},t\right) - \Theta_{a,c}\left(-\frac{L}{2},t\right)\right]\right\ra ,
 \end{split}
  \label{V1}
 \ee
We then compute the time derivative of the dual field
 \be
 \begin{split}
 \dot\Theta_{a,c}(x,t) &= i[H,\Theta_{a,c}(x,t)]\\
 & = -i\pi u\int^x_{-\frac{L}{2}}\rd y \sin\left( \sqrt{8\pi}\Phi_{a,c}(y,t) \right). 
 \end{split}
 \label{dot}
 \ee
Combining Eqs.~\fr{V1} and~\fr{dot}, we derive the expression for the voltage drop across the system: 
\be
V = Lu^2 \sum_{i=1,2} \int \rd x \int_{-\infty}^{\infty} \rd t\ \sin(\pi J t) \la {\cal H}_U(x,t) {\cal H}_U (0,0) \ra. \label{V} 
%\nn %&& \times \left\la \cos\Big(\sqrt{8\pi} \Phi_{i,c}(x,t)\Big)\cos\Big(\sqrt{8\pi}\Phi_{i,c}(0,0)\Big)\right\ra. 
\ee
To zeroth order in the coupling $u$, the finite temperature $T$ correlation function follows from Eq.~\fr{2pf}
\bea
&&  \left\la \cos\Big(\sqrt{8\pi} \Phi_{i,c}(x,t)\Big)\cos\Big(\sqrt{8\pi}\Phi_{i,c}(0,0)\Big)\right\ra \nn
&& = \left[ \frac{\pi T}{\sinh\Big( \pi T (t - x - i 0) \Big)} \frac{\pi T}{\sinh\Big(\pi T (t+x - i0)\Big)} \right]^{2K_c} \!\!\!\!\!\!\!\!\!\!\!. \nonumber
\eea
Inserting this expression into \fr{V} we find the resistivity
\be
\varrho(T) \sim u^2 T^{-3 + 4K_c}. \label{highT}
\ee
In this simple manner we reproduce the well-known result of Ref.~\cite{GiamarchiPRB91}. Linear resistivity is recovered in the limit $K_c \to 1$. Expanding  Eq.~(\ref{V}) further in the coupling $u$, we arrive at the scaling law
\be
\varrho(T) =  T f\left(\frac{T}{T^\ast}\right), \label{rho} 
\ee
where $T^\ast \sim u^{1/2(1-K_c)}$, and the scaling function satisfies $f(x) \sim x^{-4(1-K_c)}$ for $x\gg1$ and $f(x) \sim \exp(\alpha/x)$ for $x \ll 1$, where $\alpha$ is a constant of order one. The low-temperature limit is determined by the presence of a gap. From Eq.~\fr{rho} we conclude that the contribution to the conductivity has a maximum around $T^\ast$.

%%%%%%%%%%%%%
{\it Quadratic resistivity for $T \ll T^\ast$.---}
%%%%%%%%%%%%%
Let us now consider the low-$T$ PS phase. At sufficiently low $T$, the PS has opened around the antinodes, leading to the well-known nodal pockets. In this limit, there are no single electron states in the vicinity of the U-surface, and resistivity is dominated by conventional scattering occurring at the pockets. This produces $\varrho(T) \propto T^2$ via the usual argument for a Landau-Fermi liquid. Namely, the number of scattering channels varies as $T^2$ when energy and momentum conservation is imposed, leading to $\varrho(T) \propto T^2$. 

%%%%%%%%%%%%%
{\it The transition region, $T \lesssim T^\ast$.---}
%%%%%%%%%%%%%
In the low-$T$ PS phase, antinodal excitations are gapped and the nodal pockets give $\varrho(T) \propto T^2$. On the other hand, within the `strange metal' $T > T^\ast$ the resistivity is dominated by U-scattering, leading to $\varrho(T) \propto T$. Our explanation naturally leads to a crossover between these, as follows.

There exist eight hotspots where the Fermi surface and the U-surface coincide. However, much of the Fermi surface is close to the U-surface and so, at finite temperature, a significant density of electrons resides on it. These electrons experience extremely strong U-scattering, as explained above, and dominate the resistivity, leading to $\varrho(T) \propto T$ at high $T$. As the temperature is lowered, the number of electrons that undergo U-scattering is reduced. At the cross over to the PS phase at $T=T^\ast$, a gap opens about the antinodes that increases with decreasing temperature. Eventually this gap \textit{encompasses the U-surface in the antinodal regions}, and no electronic states exist in which U-scattering can occur. At this point, the resistivity is dominated by the pockets, leading to $\varrho(T) \propto T^2$. Between these two limits, a crossover occurs. 

More formally this can be extracted from Eqs.~(\ref{sigma}) and~(\ref{rho}). At a crude level, we can use an extrapolation formula that combines the antinodal contribution to the conductivity at low $T$, $\s_\text{antinodes}(T) \propto T^{-1}\exp(-\alpha T^{\ast}/T)$, with the nodal contribution, $\s_\text{nodes}(T) \propto 1/T^{2}$,  
\be
\varrho(T) = A_1 T\left[ \exp\left(-\frac{\alpha T^{\ast}}{T}\right) + \frac{BT^\ast}{T} \right]^{-1}. \label{scaling}
\ee
Here $A_1$, $B$ and $\alpha$ are constants. It is worth emphasizing that this behavior is very unusual. Upon cooling, quasiparticle excitations within much of the Brillouin zone become gapped. It would then be natural to expect that the resistivity \textit{increases}, yet, on the contrary, the resistivity drops from $\varrho(T) \propto T$ to $\varrho(T) \propto T^2$. 

\begin{figure}
\includegraphics[width=0.37\textwidth]{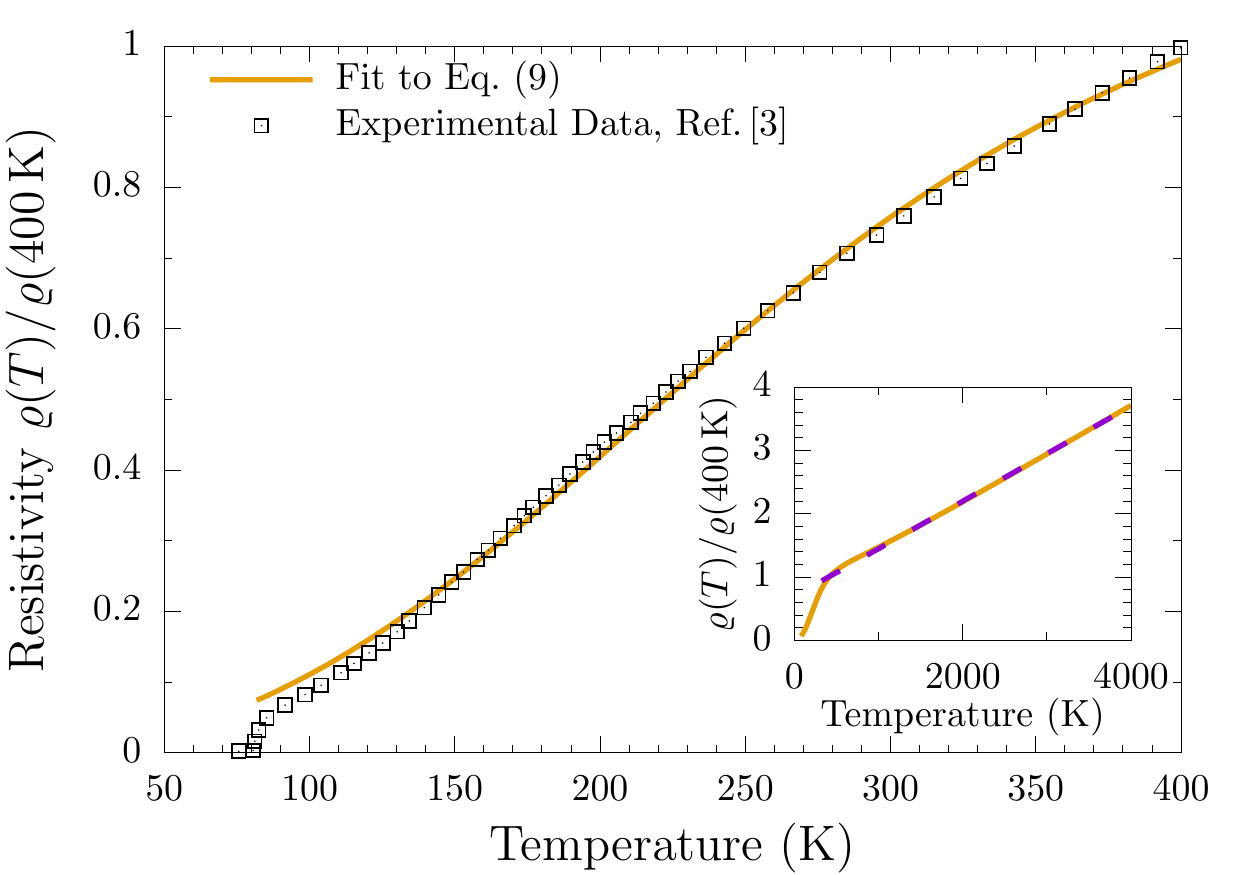}
\caption{The crude scaling form~\fr{scaling} fit to experimental data, extracted from Fig.~2 of Ref.~\cite{BarisicPNAS13}, for the resistivity as a function of temperature in HgBa$_2$CuO$_{4+\delta}$ at $p=11\%$ doping. Here $A_1/\varrho(400\,\text{K}) \approx 8\times10^{-4}$\,K$^{-1}$, $B \approx 0.25$, $\alpha\approx 2.8$, and we take $T^\ast = 280$ K. (Inset) The fit to Eq.~\fr{scaling} plotted to higher temperatures, showing linear in $T$ behavior over many decades (dashed line $\propto T$ is a guide to the eye).}
\label{Fig:Comparison}
\end{figure}

Lastly we compare our model to the detailed experiments of \textcite{BarisicPNAS13}, who analyzed the temperature dependence of the resistivity of the normal state in a series of under- and overdoped  cuprates. In Fig.~\ref{Fig:Comparison} we compare the resistivity in HgBa$_2$CuO$_{4+\delta}$ with a fit to the crude scaling form~\fr{scaling}. We see that there is excellent agreement.

\begin{figure}
\begin{tabular}{ll}
\includegraphics[width=0.21\textwidth]{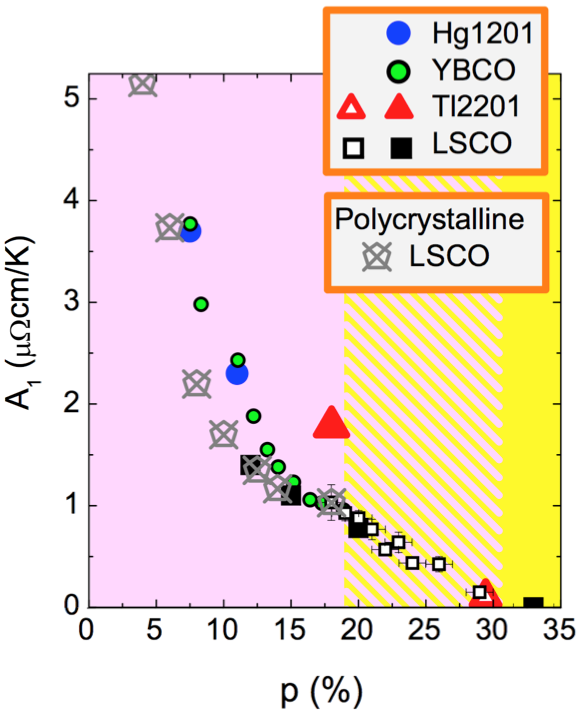} & % 0.18225, 0.225, 2025
\includegraphics[width=0.217\textwidth]{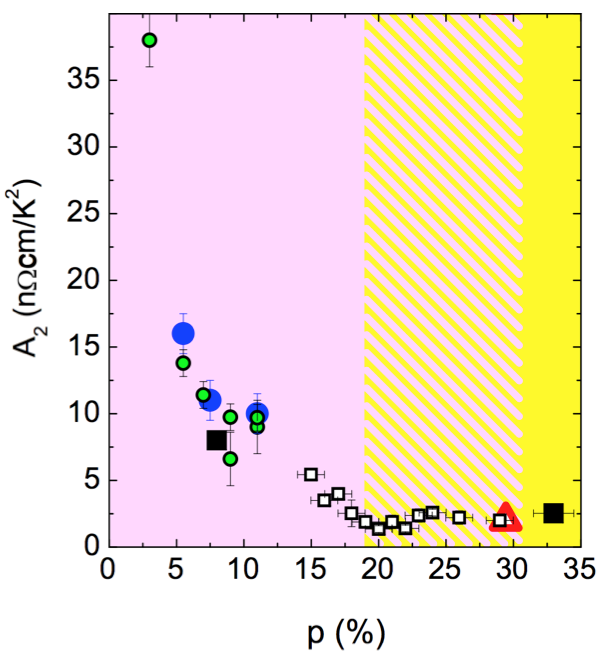} % 0.18954, 0.234, 0.2106
\end{tabular}
\caption{The doping, $p$, dependence of coefficients $A_1, A_2$ of the resistivity $\varrho(T) = A_1(p) T$ [$\varrho(T) = A_2(p) T^2$] at high [low] temperatures in a number of cuprate superconductors. Figure adapted from Ref.~\cite{BarisicPNAS13}, where further details can be found.}
\label{ExperimentalDataPlot}
\end{figure}

\textcite{BarisicPNAS13} found that the results for different cuprates could be coalesced into a single curve when they normalized their hole densities to holes per Cu$_4$O$_4$ plaquette. In the PS phase, $p < 0.18$, they fit their results to a resistivity of the form $\varrho(T) = A_1(p)T$ at high temperatures and $\varrho(T) = A_2(p) T^2$ at low-temperatures; the doping-dependent coefficients $A_1(p)$, $A_2(p)$ are displayed in Fig.~\ref{ExperimentalDataPlot}. Both coefficients increase dramatically as $p$ decreases within the PS phase. This behavior is consistent with Eq.~\fr{highT}, as the PS state is approaching the MI -- a state with a substantial single particle energy gap. The screening at low energies is reduced, leading to an increase in the U-interaction, $u$ in Eq.~\fr{umklapp}. The length of the antinodal gapped region is also enhanced. $A_1(p)$ is finite but small in the overdoped region $p > 0.20$, consistent with the one-loop FRG calculations of Ref.\,\cite{BuhmannPRB13a}. 

{\it The Hall angle.---} Our scenario can also explain an old mystery, that of the two scattering rates observed in the longitudinal and Hall conductivities~\cite{HarrisPRL95,*TerasakiPRB95,*KimuraPRB96,*BarisicArxiv15}. The magnetic field in transport coefficients scales with $T^2$, which we suggest is related to the scattering rate of the nodal quasiparticles, while the longitudinal conductivity suggests scattering rate linear in $T$. Such a separation of scattering rates naturally follows from our theory, where nodal and antinodal fermions interact weakly with each other. The nodal pockets are conventional Fermi liquids, giving rise to the $T^2$ transport coefficient in a magnetic field. Instead, the antinodal regions can be mapped onto an effective ladder model with \textit{with zero interchain tunneling}~\cite{TsvelikPRB17}. The absence of interchain tunneling is implied by the exact degeneracy of the two bands and explains why these regions---which give $T$-linear contributions to the longitudinal resistivity---are unaffected by the magnetic field. In order for the magnetic field to influence the transport of antinodal quasiparticles, the quasiparticles must be able to move around a closed path that encloses magnetic flux. When interchain tunneling is absent, no such paths can be formed and hence the antinodal regions do not contribute to the Hall conductance. In~\cite{supplemental} we use Ong's geometric analysis of the Hall effect in 2D metals~\cite{ong1991geometric} to explain the anomalous behavior of the Hall constant in the PS phase.

%%%%%%%%%%%%%
{\it Conclusions.---}
%%%%%%%%%%%%%
In this work we have proposed a simple mechanism for the origin of the linear resistivity in the normal phase of high temperature cuprate superconductors. As with a multitude of other works on the normal phase, including the enigmatic PS, this mechanism places U-scattering at its heart. Elastic U-scattering within the `strange metal' phase of the cuprates leads to $\varrho(T) \propto T$. As U-scattering flows to strong coupling under the FRG within microscopic models of the cuprates in the normal phase~\cite{HonerkampPRB01}, it is reasonable to expect that this is the dominant contribution to the resistivity. Decreasing the temperature, the PS opens and restricts the available U-scattering channels, leading to a crossover between the linear resistivity of the `strange metal' and the traditional $T^2$ form of the scattering from the Fermi pockets. 

The analogy between the physics of ladders and under doped cuprate superconductors explains phenomena beyond the transport discussed here~\cite{DagottoScience96,*DagottoRepProgPhys99}. The excitation spectrum of the fermionic ladders contains gapped collective modes that match those observed in the cuprates. The most obvious one is the spin $S=1$ magnon~\cite{DagottoPRB92,*TsunetsuguPRB94,*TroyerPRB96,*WhitePRB97}, while another is the cooperon -- a gapped bound state of two electrons (holes)~\cite{LinPRB98,KonikPRB01,KonikPRB00,*KonikPRL06,*KonikPRB10}.  Ongoing work suggests that the interaction of the cooperon with nodal quasiparticle pairs leads to a kink appearing within the spectrum of nodal quasiparticles~\cite{TsvelikUnpublished}.

The path towards detailed microscopic calculations is clear. The wave packet approach to interacting fermions, introduced by Ossadnik~\cite{OssadnikArxiv16}, allows one to map the full 2D problem to that of 1D (effective) two-leg ladders. This then opens the door to calculations that use the non-perturbative tools available for ladder systems, such as bosonization and refermionization~\cite{BalentsPRB96,LinPRB98,ControzziPRB05,*JaefariPRB12,*RobinsonPRB12,*JamesArxiv17,TsvelikPRB17} and integrability~\cite{KonikPRB01,KonikPRB02,*EsslerFieldsToStrings12}, to further study the contribution of U-scattering to the resistivity.

\acknowledgments
{\it Acknowledgments.---}
We thank Ivan Bo\v{z}ovi\'c, Peter Johnson, and Dirk van der Marel for useful and interesting discussions. We thank Subir Sachdev for useful correspondence. This work was supported by the Condensed Matter and Materials Science Division of Brookhaven National Laboratory, under the auspices of the U.S. DOE, Office of Basic Energy Sciences, Contract No. DE-SC0012704. N.J.R. has received funding from the European Union's Horizon 2020 research and innovation programme under grant agreement No 745944.

%%%%%%%%%%%%%%%%%%%%%%%%%%%%%%%%%%%%%%%%%%%%%%%%%%%%%
\begin{center}
{\bf Supplemental Material for ``Umklapp scattering as the origin of $T$-linear resistivity in the normal state of high-$T_c$ cuprate superconductors''}
\end{center}
%%%%%%%%%%%%%%%%%%%%%%%%%%%%%%%%%%%%%%%%%%%%%%%%%%%%%
\setcounter{equation}{0}
\setcounter{figure}{0}
\setcounter{table}{0}
\makeatletter
\renewcommand{\theequation}{S\arabic{equation}}
\renewcommand{\thefigure}{S\arabic{figure}}

%%%%%%%%%%%%%%
\subsection{S1. A phase space argument for linear-in-$T$ resistivity generated by umklapp scattering in 1D} 
%%%%%%%%%%%%%%

\begin{figure}[h]
\includegraphics[width=0.3\textwidth]{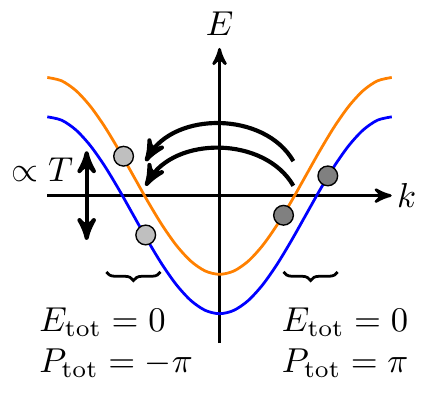}
\caption{Schematic of the phase space argument for umklapp scattering generating linear-in-$T$ resistivity. Two right-moving electrons (dark grey points) umklapp scatter to two left-moving electrons (light grey points) at half-filling and finite temperature.}
\label{Fig:phasespace}
\end{figure}

Here we present a simple phase space argument for linear-in-$T$ resistivity generated by umklapp (U) scattering in one spatial dimension. This argument is superseded by the detailed calculation presented in the main text, but is useful for illustration and captures the essential physics. Similar arguments have been constructed previously, see e.g.~\ocite{BatisticPRB93}.

Let us consider the U-scattering of electrons in a half-filled band at finite temperature. This is illustrated schematically in Fig.~\ref{Fig:phasespace} for a process where two right-moving electrons (dark grey spots) are scattered to two left-moving electrons (light grey spots). As usual, momentum changes by $|\Delta P| = 4k_F = 2\pi$ and the scattering event is a zero-energy process, $E_\text{tot,init} = 0 = E_\text{tot,fin}$.  

A Boltzmann-like argument tells us that the resistivity generated by such processes, which dissipate momentum (see also the discussion of the next section), will scale with the number of scattering channels. This can easily be estimated: we call this the phase space argument. 

Consider fixing the momentum (and hence energy) of one of the right-moving particles: this completely fixes the momentum (and energy) of the other right-moving particle through $E_\text{tot} = 0$ and $P_\text{tot} = \pi$. These are then scattered to a pair of left-moving particles: these must be symmetrically distributed about the left Fermi point in order that energy is conserved, $E_\text{tot} = 0$, and momentum is changed by $2\pi$, $P_\text{tot} = -\pi$. The number of such allowed scattering channels is then controlled by the temperature $T$: there must be a thermal population of hole states into which one can scatter one of the electrons. The energy range over which hole states are available is linear in $T$ (shown schematically in Fig.~\ref{Fig:phasespace}), and hence the number of scattering channels grows linearly with $T$, thus $\rho(T) \propto T$. 

%%%%%%%%%%%%%%
\subsection{S2. Contribution of umklapp scattering to resistivity in 1D quantum systems} 
%%%%%%%%%%%%%%

The issue of transport in one-dimensional (1D) quantum systems, and whether or not it is ballistic, is a thorny one. In particular, whether U-scattering in a clean system generates non-zero resistivity has received much attention (see, e.g., Refs.~\cite{RoschPRL00} and references therein). The crux of the issue, as realized by \textcite{RoschPRL00}, is whether there are additional conservation laws beyond energy and momentum, that can lead to ballistic transport (i.e., a non-zero Drude weight). Let us briefly recount the main argument of~\cite{RoschPRL00} and discuss its implications in the scenario that we consider. 

Consider a single band Luttinger liquid and the influence of U-scattering at \textit{incommensurate filling}
\bea
H &=& H_0 + H_U, \label{H} \\
H_0 &=& \frac{1}{2} \int \rd x\, \sum_{\nu = s,c} \frac{v_{\nu}}{2} \Big( K_{\nu} \big( \p_x \theta_\nu )^2 + K_{\nu}^{-1} (\p_x \phi_{\nu})^2 \Big),\nn
H_U &=& \sum_{n,m} H^{n,m}_U, \nonumber
\eea
where $v_{\nu}$ ($K_\nu$) is the velocity (Luttinger parameter) for spin ($\nu=s$) and charge ($\nu=c$) bosons. $H_U^{n,m}$ describes the U-processes that scatter $n$ right-movers to $n$ left-movers (and vice versa) whilst transferring momentum $2m\pi \equiv mG$ to/from the lattice. In bosonized form, the umklapp term $H_U^{n,m}$  reads
\be
H^U_{n,m} \approx \frac{g_U^{n,m,n_s}}{(2\pi\alpha)^n} \int \rd x\, e^{i\Delta k_{n,m} x} e^{i\sqrt{2\pi}(n \phi_c + n_s \phi_s)} + {\rm H.c.}, \nonumber
\ee
where $\alpha$ is a short-distance cut-off of order the lattice spacing. Here we consider a U-scattering event in which $\Delta k_{n,m} = 2n k_F - mG$ is the momentum transferred and $n_s/2$ the total spin transferred in the scattering event. $k_F$ is the Fermi wave vector. 

For incommensurate filling U-terms are irrelevant in the renormalization group sense. However, the presence of such terms in the theory can determine the low-frequency conductivity, as these terms induce decay of the modes close to the Fermi surface. In this sense, these operators are ``dangerously irrelevant''. One key point in determining the low-frequency behavior, as realized in Ref.~\cite{RoschPRL00}, is the presence of an additional operator $O$ in the theory that commutes with \textit{any single U-term}: $[ H_U^{n,m} , O] = 0$. This operator is formed from $J_0$, the difference in numbers of right- and left-moving fermions (i.e., the current), and the translation operator $P_T$:
\bea
O &=& \Delta k_{n,m} J_0 + 2 n P_T,  \label{opO} \\
J_0 &=& \sum_\s \int \rd x \, \Big( \psi\dg_{R,\s} \psi_{R,\s} - \psi\dg_{L,\s}\psi_{L,\s}\Big), \nn
P_T &=& \sum_\s \int \rd x \, \Big( \psi\dg_{R,\s} (-i\p_x) \psi_{R,\s} + \psi\dg_{L,\s} (-i\p_x) \psi_{L,\s}\Big), \nonumber
\eea
where $\psi_{R/L,\s}$ are right/left moving spin-$\s$ fermion fields related to the bosonic fields in~\fr{H} by the usual bosonization identities~\cite{GiamarchiBook}.

When restricting attention to a model that contains a single $H_{U}^{n,m}$ term, the authors show that the presence of such an operator~\fr{opO} and the corresponding conservation law implies infinite conductivity. That is, a single U-term in a Luttinger liquid does not induce finite resistivity. This is shown explicitly within the memory matrix approach. Although the current $J_0$ does not commute with the umklapp term, it has a finite overlap with the conserved operator $O$ and this is sufficient to have ballistic transport. Instead, at incommensurate fillings, there must be at least two independent U-terms to induce finite resistivity. Similar issues have been encountered in higher dimensions, for one such example in two dimensions consult Ref.~\cite{PatelPRB14}. 

Let us now turn our attention to the scenario envisaged within the main body of our manuscript. There we examine an \textit{incommensurately filled} 2D system where strong interactions lead to a deformation of the Fermi surface that increases nesting. This involves the net transfer of electrons from the antinodal to the nodal regions, leading to the well-known pocket in the nodal region and the anitnodal region coinciding with the U-surface. The deformed Fermi surface is shown as the red curve in Fig.~2 of the manuscript. The antinodal regions that run along the U-surface are quasi-1D and are mapped to effective ladder models~\cite{TsvelikPRB17}. Crucially, these effective ladders have \textit{commensurate filling}, and the argument of~\textcite{RoschPRL00} fails to apply; the U-process $H_U^{2,2}$ that is considered has momentum transfer $\Delta k_{2,2} = 0$ and hence the current $J_0$ does not have an overlap with any conserved operator.

This can be seen clearly by reformulating the U-term in Eq.~(2) of the manuscript. As is usual with bosonization, such a term can be represented in several different forms. For example, in Ref.~\cite{TsvelikPRB17} the spin sector was represented in terms of fermions, whilst the charge sector was bosonic. It can also be convenient to fermionize the charge sector; we give both of these formulations here. According to Eq.~(10) of Ref.~\cite{TsvelikPRB17}, the U-term can be written as
\be
{\cal H}_U = u \cos(\sqrt{4\pi}\Phi_{c}) {\cal M} = \tilde u {\cal M} (r^+l +l^+r), \label{umklappRewrite}
\ee
with $u \sim \tilde u$ the U-interaction strength, $\Phi_c$ the total charge bosonic field in the low-energy description of the ladder, $(r,l)$ are right/left moving fermions that carry charge $e$, but neither spin nor leg index. The operator ${\cal M}$ has na\"ive scaling dimension one and includes all other fields of the theory (see~\cite{TsvelikPRB17} for an explicit form). In term of the charge fermions $r,l$, the current is $J = 4(r^+r - l^+l) \sim \partial_x\Theta_c $ and a direct calculation shows that $[J, \int \rd x\,{\cal H}_U] \neq 0$.

The idea that the combination of commensurate filling and dynamical gap generation is special is supported by exact results extracted from integrable models. The existence of ballistic transport is tantamount to the existence of a nonzero Drude weight $D$, defined as the prefactor of the delta-function in the a.c. conductivity, $\s(\omega) = D(T) \delta(\omega) + \s_{\rm reg}$. The Drude weight can be extracted from the excitation spectrum (see, e.g., Ref.~\cite{CastellaPRL95}). In particular,~~\textcite{ZotosPRL99} derived an expression for $D_s$, the Drude weight for spin transport in the spin-1/2 anisotropic antiferromagnet, the XXZ model, using the thermodynamic Bethe ansatz. In the limit of weak bare interactions, this describes the charge sector of the Hubbard model. In the region of the phase diagram in which we are interested, where the charge spectrum is gapped, it was shown that $D(T)=0$ at all temperatures. 
 
While the results derived from the thermodynamic Bethe ansatz were somewhat controversial, more recently the formulae for $D(T)$ was rederived using by Benz and collaborators~\cite{BenzJPSJS05} using advanced methods for finite-size systems, which leads to the same result in the gapped case~\cite{sakaiUnpublished} and confirms the absence of ballistic transport. Very recently, there have been significant advances in computing Drude weights in integrable models using the method of generalized hydrodynamics, see Refs.~\cite{BulchandaniArxiv17,IlievskiPRL17,IlievskiPRB17,DoyonArxiv17} (see in particular Ref.~\cite{IlievskiPRB17} for the one-dimensional Hubbard model).

%%%%%%%%%%%%%%
\subsection{S3. The Hall constant} 
%%%%%%%%%%%%%%

Let us recall the geometric approach to the Hall effect in two dimensions, as devised by N.~P.~Ong in Ref.~[\onlinecite{ong1991geometric}]. There it was shown that the key quantity to consider is the so-called ``scattering path length'' $\bm{l}(\bm{k})$:
\be
\bm{l}(\bm{k}) = \bm{v}(\bm{k}) \tau_{\bm{k}}, 
\ee
where $\bm{v}(\bm{k})$ is the group velocity and $\tau_{\bm{k}}$ is the relaxation time of an electron with momentum $\bm{k}$. The Hall constant is then determined by the area (in the $\bm{l}_x-\bm{l}_y$ plane) swept out by the scattering path length as the momentum $\bm{k}$ is moved around the Fermi surface. This area $A_{\bm l}$ may be computed through (see \cite{ong1991geometric} for further details)
\be
A_{\bm l} = \hat{\bm z} \cdot \int \rd \bm{l}\times\bm{l}/2. 
\ee

Let us now consider this framework applied to the PS phase of the cuprates, as described by the two fluid model. We begin with the high temperature `strange metal' phase, $T > T^\ast$. The Fermi velocity $\bm{v}_F \equiv \bm{v}(\bm{k}_F)$ is constant along the straight sections of the YRZ Fermi surface that connect the ends of the arcs to the antinodes. As a result, these sections do not contribute to the area swept out by the scattering path length (see, for example, Fig. 3 of~[\onlinecite{ong1991geometric}]), and instead one only has contributions from the nodal arcs. On the other hand, at low temperature $T < T^\ast$ the straight sections become gapped, leading once again to only the four arcs contributing to the Hall constant. Hence in both temperature regimes, $T > T^\ast$ and $T<T^\ast$, the doped hole density controls the number of carriers observed in the Hall effect. This is indeed the result observed in experiments (see, e.g., the discussion of the review~\cite{hussey2008phenomenology}). 

\bibliography{bib}

\end{document}